\documentclass[11pt]{article}
\usepackage{setspace,braket}
\usepackage{amsmath, amssymb, bm, yfonts}
\usepackage{etoolbox}
\usepackage{jheppub}
    \makeatletter
    \patchcmd{\maketitle}{\@fpheader}{}{}{}
\makeatother
\usepackage{booktabs}
\usepackage{multirow}
\usepackage{longtable}
\usepackage{moreverb,url}

\usepackage{listings}
\usepackage{balance}
\usepackage{xcolor}
\def\be{\begin{equation}}
\def\ee{\end{equation}}

\def\be{\begin{equation}}
\def\ee{\end{equation}}

\def\bg{\bar{g}}

\def\beq{\begin{eqnarray}}\def\eeq{\end{eqnarray}}
\def\ba#1\ea{\begin{align}#1\end{align}}
\def\bg#1\eg{\begin{gather}#1\end{gather}}
\def\bm#1\em{\begin{multline}#1\end{multline}}
\def\bmd#1\emd{\begin{multlined}#1\end{multlined}}

\def\({\left(}
\def\){\right)}
\def\[{\left[}
\def\]{\right]}

\begin{document}

\title{EasyABM: a lightweight and easy to use heterogeneous agent-based modelling tool written in Julia.}
\author[a]{Renu Solanki}
\author[b]{, Monisha Khanna}
\author[a]{, Shailly Anand}
\author[a]{, Anita Gulati}
\author[c]{, Prateek Kumar}
\author[b]{, Munendra Kumar}
\author[]{, Dushyant Kumar}
\affiliation[a]{Deen Dayal Upadhyaya College, University of Delhi, Sector 3, Dwarka, New Delhi, India - 110078}
\affiliation[b]{Acharya Narendra Dev College, University of Delhi, Govindpuri, Kalkaji, New Delhi, India - 110019}
\affiliation[c]{Department of Zoology, University of Allahabad, Prayagraj, Uttar Pradesh, India -211002}

\begin{abstract}{
Agent based modelling is a computational approach that aims to understand the behaviour of complex systems through simplified interactions of programmable objects in computer memory called agents. Agent based models (ABMs) are predominantly used in fields of biology, ecology, social sciences and economics where the systems of interest often consist of several interacting entities. In this work, we present a Julia package EasyABM.jl for simplifying the process of studying agent based models. EasyABM.jl provides an intuitive and easy to understand functional approach for building and analysing agent based models.}
\end{abstract}
\maketitle
\onehalfspace

\section{Introduction}
Agent based modelling is a computataional paradigm in which autonomous interacting entities called agents are evolved in time according to predefined rules with an aim to understand emergent phenomena in complex systems. Agent based modelling has a diverse range of applications in economics, social behaviour, biology, epidemiology and ecology. In particular, agent based models (ABMs) have been used to study pandemics~\cite{Nic2020, Hinch2021, Kerr2021}, population dynamics~\cite{SM2014, Jang2016}, biofilms~\cite{Lardon2011}, tumor growth~\cite{Jan2016}, financial markets~\cite{Sam2020, Feng2012} as well as societal phenomena such as
urbanisation~\cite{Mau2021}, traffic flows~\cite{Hager2015} and migration~\cite{Anna2016, Jule2018}.

 Depending on the application, an agent in an ABM can represent an organism, institution, a physical object or an abstract entity. There is generally no central control system in ABMs and agents act independently according to the rules of the model. Surprisingly, even simple rules of dynamics can lead to interesting emergent phenomena which is a key driving principle behing the usefulness of ABMs in understanding complex systems. 

Several frameworks and tools have been developed for agent based modelling. Some of the general purpose ABM frameworks include NetLogo \cite{Wilensky1999}, Mesa~\cite{DMas2015}, Agents~\cite{George2022},  MASON~\cite{SLuk2005}, Swarm~\cite{HIba2013} and Repast~\cite{MJ2013}. All these frameworks use different programming languages and follow different design principles. Until now, Agents.jl has been the only ABM framework written in Julia. Our objective in developing EasyABM.jl has been to provide Julia users interested in agents based modelling with an alternate tool. EasyABM.jl follows a completely different design principle from Agents.jl. The API of EasyABM takes a function based approach where for each modelling task from creating agents and model, to running model and creating visualisation, there is a function to accomplish the task. Agents.jl is similar except that one needs to use a struct for creating agents. Even though it leads to a more performant code, the idea of a struct may not be easily comprehensible for people with only a basic background in programming. Another aspect of Agents.jl which, in our opinion, makes it somewhat complex to understand for new users is two different types of step functions (\verb+agent_step!+ and \verb+model_step!+) that can be used for simulation, compared to only one step function (\verb+step_rule!+) in EasyABM. Visualisation is another area where EasyABM takes relatively more intuitive approach compared to Agents.jl. In EasyABM the graphics properties are associated with agents at the time of definition of agents or at model initialisation step and later can be manipulated during model run. Visualisation can then by done by a single function call whereas in Agents.jl it takes more effort.

\begin{figure}[tbh]
\centering
\includegraphics[scale = 0.12]{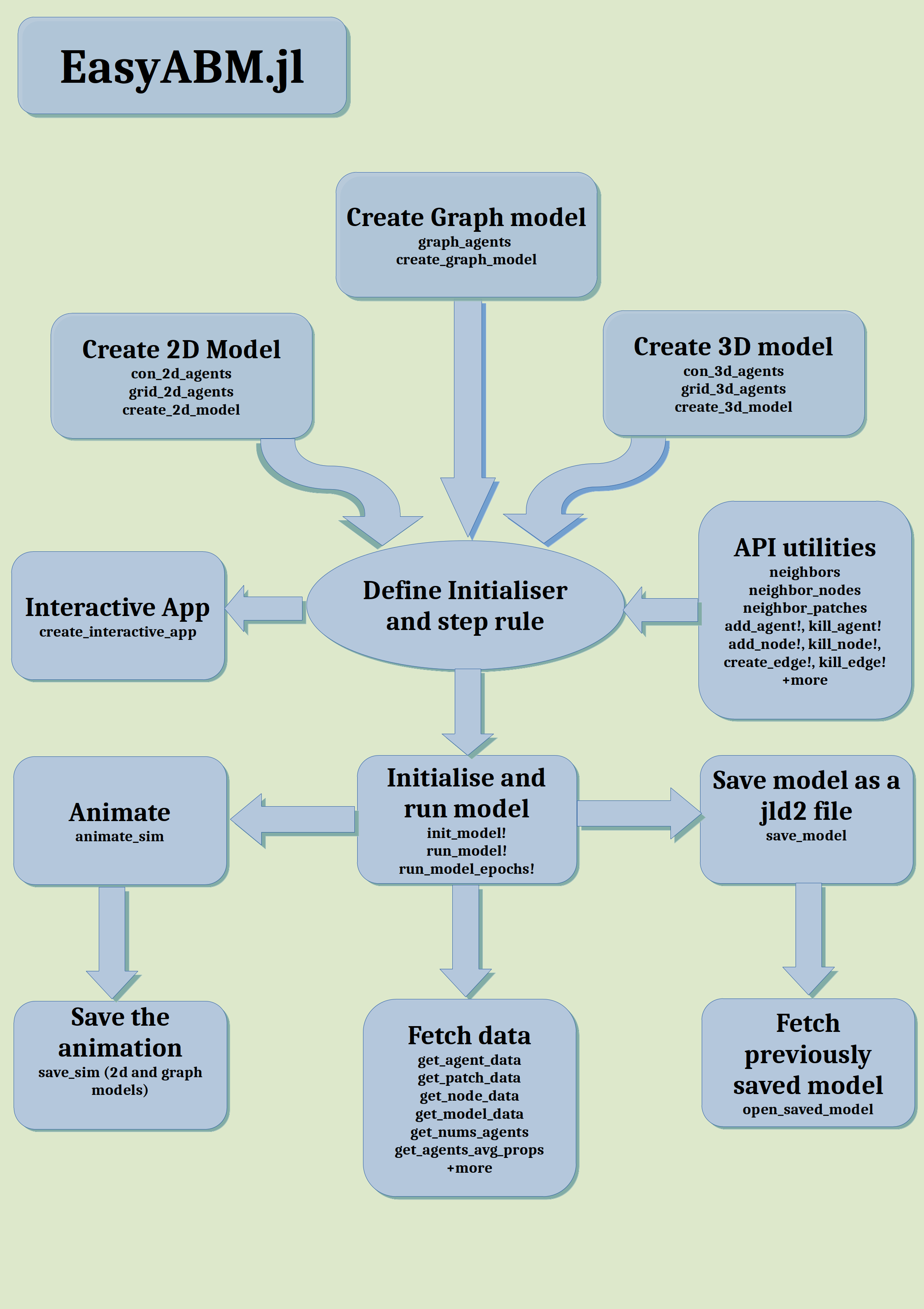}
\caption{\label{fig:workflow} Agents based modelling workflow with EasyABM.}
\end{figure}

The figure~\ref{fig:workflow} gives a diagrammatic representation of agents based modelling workflow with EasyABM. In the next section we explain this workflow with some examples of agents based models.

\section{Agent based modelling with EasyABM}
EasyABM.jl allows users to create and study models in 2D, 3D and graph spaces. In the 2D and 3D models there can be two types of agents - grid agents which can move only in discrete steps and continuous agents which can move continuously. Moreover, both in 2D and 3D models the space is divided into unit blocks called patches where each patch can have its own properties like agents. This feature of EasyABM.jl has been inspired by Netlogo\cite{Wilensky1999}. Nodes and edges in a graph space can also be assigned properties similar to agents. A graph space can be chosen to be fully dynamic in which nodes and edges can be added or removed during time evolution.

In the present section we explain the workflow of EasyABM, as outlined in Figure~\ref{fig:workflow}, through flocking model in 2D, Schelling's segregation model in 3D and Ising model on a nearest neighbor graph for graph spaces. For more details on these and many other models in EasyABM we refer to the examples in the online documentation. 

\subsection{2D models}
EasyABM uses functional approach for creation of agents and models. The code in listing~\ref{lst:flockingdefine} uses the function \verb+con_2d_agents+
 to create 200 continuous space 2D agents with properties shape, pos, vel and orientation and then creates the model using \verb+create_2d_model+ function. For a grid based 2D model one needs grid agents which can be created using \verb+grid_2d_agents+ function.

\begin{lstlisting}[caption = {Flocking in 2D - Defining agents and the model.}, label={lst:flockingdefine}]
using EasyABM

boids = con_2d_agents(200, 
		shape = :arrow, pos = Vect(0.0,0.0), 
		vel=Vect(0.0,0.0), orientation = 0.0, 
		keeps_record_of = [:pos, :vel, 
		:orientation])

model = create_2d_model(boids,
		agents_type=Static,
		space_type = Periodic, 
		min_dis = 0.3, coh_fac = 0.05, 
		sep_fac = 0.5, dt= 0.1, 
		vis_range = 2.0, aln_fac = 0.35)
\end{lstlisting}

EasyABM recognises pos as position and vel as velocity of the agent. The property pos must be a Vect which is an inbuilt vector type of EasyABM. It is also convenient to use Vect for any other vectorial properties in the model like velocity or forces. The properties shape and orientation are also recognised by EasyABM as graphics properties of agents. The \verb+keeps_record_of+ argument is the list of properties that the agent will record during time evolution. The agents in EasyABM are fully hetrogeneous in that each agent can have different set of properties of which it can record any specified subset during model run. The list of agents, that we have named as boids for the flocking model, is supplied as the first argument in the function \verb+create_2d_model+ for creating model. The argument \verb+agents_type+ is set to Static if the number of agents remain fixed during simulation and Mortal if the agents can take birth or die. The \verb+space_type+ property is set to Periodic if the space is periodic in all directions and NPeriodic for a non-periodic space. Other arguments of the \verb+create_2d_model+ function are parameters specific to the flocking simulation:
\begin{itemize}
\item \verb+min_dis+ : The distance between boids below which they start repelling each other.
\item \verb+coh_fac+ : The proportionality constant for the cohere force.
\item \verb+sep_fac+ : The proportionality constant for the separation force.
\item \verb+aln_fac+ : The proportionality constant for the alignment force.
\item \verb+vis_range+ : The visual range of boids.
\item dt : The proportionality constant between change in position and velocity.
\end{itemize}

After creating the model, the initial properties of agents can be set through an initialiser function which is then sent as an argument to \verb+init_model!+ function provided by EasyABM.jl. The code for the initialiser function for 2D flocking model in shown in listing~\ref{lst:flockinginit}

\begin{lstlisting}[caption = {Flocking in 2D - Initialiser function.}, label={lst:flockinginit}]
function initialiser!(model)
    xdim, ydim = model.size
    for boid in model.agents
        boid.pos = Vect(rand()*xdim, rand()*ydim)
        boid.orientation = rand()*2*3.14
        boid.vel = Vect(-sin(boid.orientation), 
        		cos(boid.orientation))
    end
end

init_model!(model, initialiser = initialiser!)
\end{lstlisting}

The next step is to define rules of dynamics through a \verb+step_rule!+ function and run the model. The \verb+step_rule!+  function for the flocking model is shown in listing~\ref{lst:flockingstep}.  

\begin{lstlisting}[ caption = {Defining step rule and running the model.}, label={lst:flockingstep}]
const ep = 0.00001

function step_rule!(model)
    dt = model.parameters.dt
    min_dis = model.parameters.min_dis
    aln_fac = model.parameters.aln_fac
    coh_fac = model.parameters.coh_fac
    sep_fac = model.parameters.sep_fac
    for boid in model.agents
        nbrs = euclidean_neighbors(boid, model, 
        model.parameters.vis_range)
        coh_force = Vect(0.0,0.0) 
        sep_force = Vect(0.0,0.0) 
        aln_force = Vect(0.0,0.0)
        num = 0
        for nbr in nbrs
            num+=1
            vec = nbr.pos - boid.pos
            if veclength(vec)< min_dis
                sep_force -= vec
            end
            coh_force += vec
            aln_force += nbr.vel
        end
        if num>0
        	 aln_force = (aln_force/ num - 
        	 boid.vel)*aln_fac
        	 coh_force *= coh_fac/num
        	 sep_force *= sep_fac
        	 boid.vel  += (coh_force + sep_force) + aln_force
        	 boid.vel  /= (veclength(boid.vel)+ep)
        	 boid.orientation = calculate_direction(boid.vel)
        end
        boid.pos += boid.vel*dt
    end
end

run_model!(model, steps=500, 
						step_rule = step_rule!)
\end{lstlisting}

\begin{figure}[tb]
\centering
\includegraphics[width=\columnwidth]{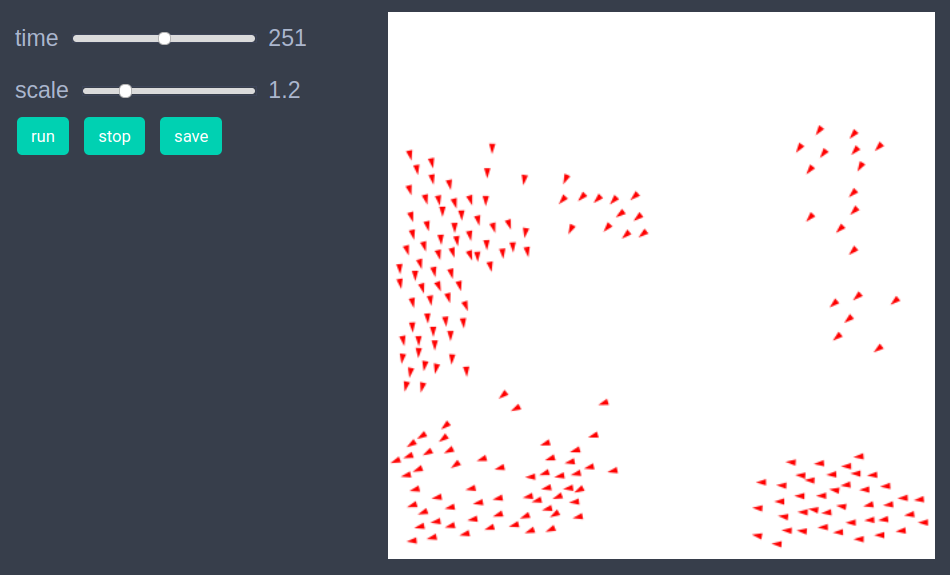}
\caption{\label{fig:flockanim} Visualisation for flocking model. The size of agents can be changed with the size slider. Save button can be used for saving the animation.}
\end{figure}

With the data collected during model run, the animation can be created with a single line of code as shown below. The output of this line of code is an interactive animation as shown in Figure~\ref{fig:flockanim}. The save button can be used to save the animation to the path mentioned in the \verb+animate_sim+ function.

\begin{lstlisting}
animate_sim(model, path = "/path/to/folder/anim.gif")
\end{lstlisting}

\begin{figure*}[tb]
\centering
\includegraphics[width=\textwidth]{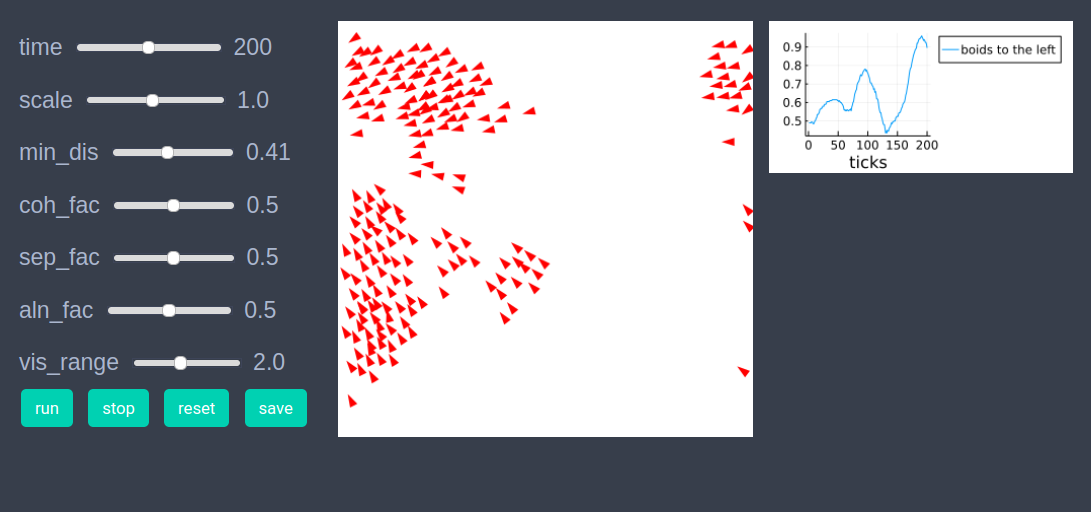}
\caption{\label{fig:flockapp}Interactive app for flocking model. Sliders can be used to change model parameters and then the model can be reset and run again. The plot on the right shows fraction of boids which are in the left half of the space.}
\end{figure*}

EasyABM also makes it very easy to create an interactive application for the model in Jupyter and similar notebook environments with WebIO installation. The following lines of code create an interactive app for the flocking model as shown in Figure~\ref{fig:flockapp}.

\begin{lstlisting}
create_interactive_app(model,
			initialiser= initialiser!,
			step_rule= step_rule!,
			model_controls=
				[(:min_dis, :s, 0.01:0.1:1.0),
				(:coh_fac, :s, 0.01:0.01:1.0),
				(:sep_fac, :s, 0.01:0.01:1.0),
				(:aln_fac, :s, 0.01:0.01:1.0),
				(:vis_range, :s, 0.5:0.5:4.0)],
			agent_plots=Dict("boids to the left"=> 
        	boid -> boid.pos[1]<model.size[1]/2),
				frames=400) 
\end{lstlisting}

Finally, EasyABM.jl API provides functions to fetch data of individual agents and also data averaged over all agents. For example the data of the agent with index 1, can be fetched as
\begin{lstlisting}
df = get_agent_data(model.agents[1], model).record
\end{lstlisting}

Similarly, the average velocity of agents at all times during model run can be obtained as 

\begin{lstlisting}
df = get_agents_avg_props(model, 
	agent -> agent.vel, 
	labels = ["average velocity"])
\end{lstlisting}

Similar functions are also available for fetching data of patches, nodes and edges. 

\subsection{3D models}
Just as for 2D models, in 3D too EasyABM.jl provides simple functions to accomplish modelling tasks. We explain its use in 3D through the 3D version of Schelling's segregation model. In the Schelling's model agents live on a grid. The code below uses the function \verb+grid_3d_agents+ to create 200 3D-grid agents with properties  color, pos and mood and then uses function \verb+create_3d_model+ to create the model. For a model requiring a continuous 3D space agents can be created using \verb+con_3d_agents+ function, while the function for creating model remains the same.

\begin{lstlisting}
using EasyABM

@enum agentsfeeling happy sad

agents = grid_3d_agents(200, pos = Vect(1,1,1), 
			color = :red, mood = happy, 
			keeps_record_of=[:pos, :mood])
			
model = create_3d_model(agents, 
			agents_type = Static, 
			space_type = NPeriodic, 
			size = (7,7,7), min_alike = 8)
\end{lstlisting}

For grid type agents the position vector is required to have integer valued coordinates. The mood of agents is an enum property which can take only two values - happy or sad. Also through \verb+keeps_record_of+ argument, we are asking agents to record their position and mood properties during time evolution. Similar to \verb+create_2d_model+ function, the function \verb+create_3d_model+ also accepts the list of agents as the first argument. Since the number of Schelling's agents remain fixed during simulation we set the argument \verb+agents_type+ to Static. We also set \verb+space_type+ property to  NPeriodic for a non-periodic 3D space. The argument \verb+min_alike+ is a model property specific to the Schelling's model and we set its value to 8.

After creating the model, the initial properties of agents can be set through an initialiser function which is then sent as an argument to \verb+init_model!+ function provided by EasyABM.jl. The code for the initialiser function for 2D flocking model in shown below.

\begin{lstlisting}
function initialiser!(model)
    for agent in model.agents
        agent.color = [:red, :green][rand(1:2)]
        x,y,z = random_empty_patch(model) 
        agent.pos = Vect(x, y, z)     
    end    
    for agent in model.agents
        nbrs = grid_neighbors(agent, model, 1)
        num_same = 0
        for nbr in nbrs
            if nbr.color == agent.color
                num_same += 1
            end
        end
        if num_same < model.parameters.min_alike
            agent.mood = sad
        end
    end
end
init_model!(model, initialiser = initialiser!)
\end{lstlisting}

The \verb+step_rule!+ function for the 3D Schelling's model is shown in listing~\ref{lst:schellingstep}. This function is then passed as an argument to the \verb+run_model!+ function to run it for required number of steps.  

\begin{lstlisting}[caption = {Schelling's segregation in 3D - step rule.}, label={lst:schellingstep}]
function step_rule!(model)
    min_alike = model.parameters.min_alike
    for agent in model.agents
        num_alike = 0
        for nbr in neighbors(agent, model,1)
            if agent.color == nbr.color
                num_alike += 1
            end
        end
        if num_alike >= min_alike
            agent.mood = happy
        else
            agent.mood = sad
            x,y,z = random_empty_patch(model) 
            agent.pos = Vect(x, y, z)
        end
    end
    return
end

run_model!(model, steps=200, 
step_rule = step_rule!)
\end{lstlisting}

EasyABM.jl uses MeshCat.jl as backend for 3D visualisations. Animation from the data collected during model run can be created with the \verb+animate_sim+ function as in case of 2D models.

\begin{figure*}[tb]
\centering
\includegraphics[width=\textwidth]{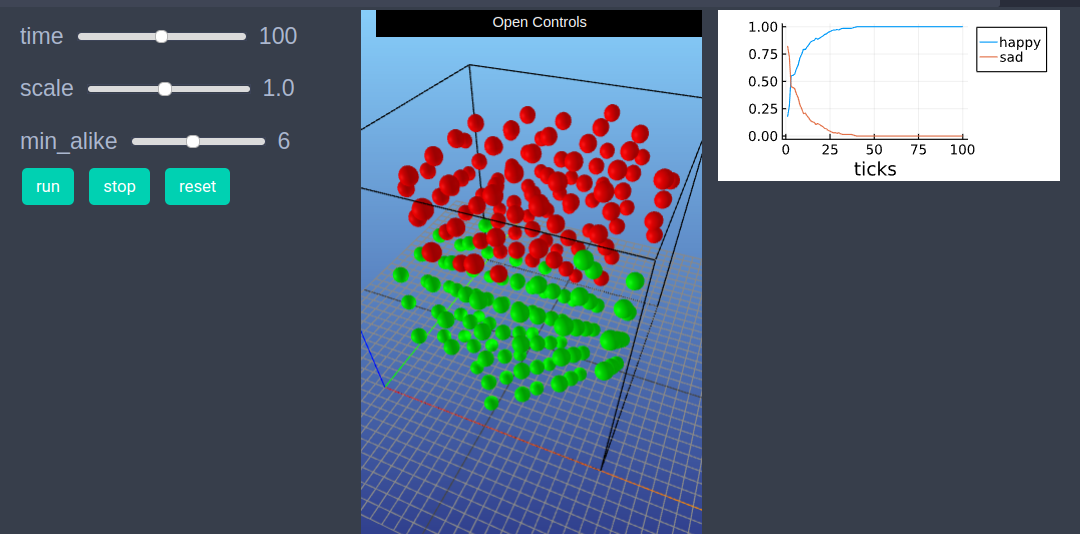}
\caption{\label{fig:schellapp}Interactive app for Schelling's 3D model. The plot on the right shows fraction of happy and sad agents in the model as the time progresses.}
\end{figure*}

Creating an interactive application also requires a single function call with intuitive and easy to understand arguments. The following function creates an interactive app for the Schelling's 3D model as shown in Figure~\ref{fig:schellapp}.

\begin{lstlisting}
create_interactive_app(model, 
	initialiser= initialiser!,
    step_rule=step_rule!,
    model_controls=[(:min_alike, :s, 1:12)], 
    agent_plots=Dict(
    "happy"=> agent-> agent.mood == happy, 
    "sad"=> agent-> agent.mood == sad
    ),
    frames=200) 
\end{lstlisting}

\begin{figure}[tb]
\centering
\includegraphics[width=\columnwidth]{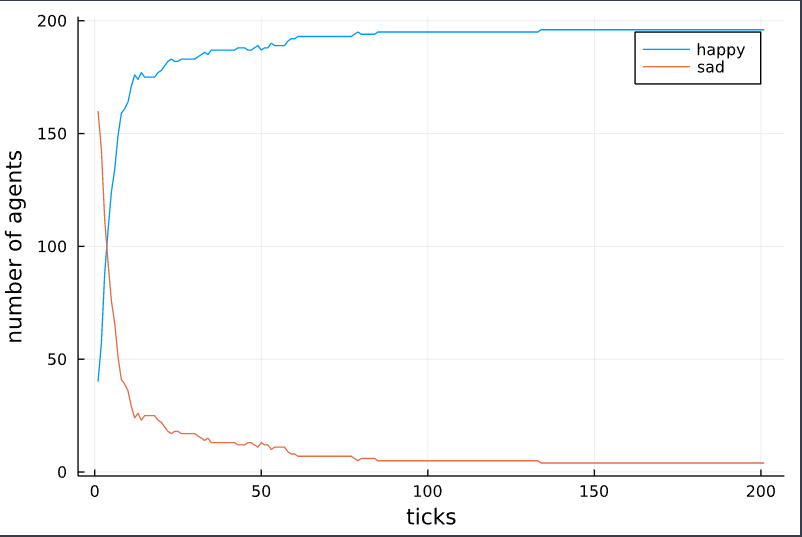}
\caption{\label{fig:schellplot}Plot of number of happy and sad agents along time evolution of Schelling's 3D model.}
\end{figure}

Data of individual agents, as well as average properties of agents can be fetched via simple function calls. The code below fetches the data of number of happy and sad agents at each time step as a dataframe and plots the result as shown in Fig~\ref{fig:schellplot}. 

\begin{lstlisting}
df = get_nums_agents(model, 
agent-> agent.mood == happy, 
agent-> agent.mood == sad,
labels=["happy","sad"], 
plot_result=true)
\end{lstlisting}

\subsection{Graph based models}

Graph spaces in EasyABM.jl can be either static or dynamic. The topology of a static graph remains fixed during model run, while that of a dynamic graph can be changed. Graphs created with other julia packages like Graphs.jl can be used in graph based models after converting them to static or dynamic type using inbuilt EasyABM functionality. In this subsection we explain the workflow for graph based models using example of Ising model on nearest neighbor graphs. In graph based models agents can be created using \verb+graph_agents+ function and then included in the model in the same way as in the case of 2D and 3D models. However, for the Ising model, we will not need any agents and would rather attach spin property to the nodes of the graph. As shown in the code below, we define the model by creating an empty graph and using \verb+create_graph_model+ function with the graph as the first argument. We set the \verb+agents_type+ to Static as the number of agents is fixed to be zero. There are three parameters \verb+temp+, \verb+coupl+ and \verb+nns+ in the model where \verb+temp+ and \verb+coupl+ are respectively the temperature and coupling parameters of the Ising model and \verb+nns+ is the number of nearest nodes that each node will have an edge with in the graph.

\begin{lstlisting}
graph = dynamic_simple_graph(0)
model = create_graph_model(graph, 
		agents_type = Static,
		temp = 2.0, coupl = 2.5, 
		nns = 5) 
\end{lstlisting}

In the code for initialisation, as shown in listing~\ref{lst:Isinginit}, we use NearestNeighbors.jl package and define an initiliser function where a nearest neighbor graph of a fixed number of nodes is created and its nodes are randomly assigned with spin and color properties, where we assign black color to nodes with spin 1, and white color to nodes with spin -1. The properties spin and color of nodes that we want to be recorded during model run are specified in the \verb+props_to_record+ argument in the \verb+init_model!+ function.

\begin{lstlisting}[caption = {Ising model on nearest neighbor graph - initialisation.}, label={lst:Isinginit}]
using NearestNeighbors

const n = 500; 

function initialiser!(model)
    vecs = rand(2, n)
    kdtree = KDTree(vecs,leafsize=4)
    flush_graph!(model)
    add_nodes!(n, model, color = :black, spin =1)
    for i in 1:n 
        model.graph.nodesprops[i].pos = (vecs[1,i], vecs[2,i]) 
        indices, _ = knn(kdtree, 
        			vecs[:,i], 
        			model.parameters.nns, true)
        for j in indices
            if j!=i
                create_edge!(i,j, model)
            end
        end
        if rand()<0.5
            model.graph.nodesprops[i].spin = 1
            model.graph.nodesprops[i].color = :black
        else
            model.graph.nodesprops[i].spin = -1
            model.graph.nodesprops[i].color = :white
        end
    end
end

init_model!(model, initialiser= initialise!, 
    props_to_record = Dict("nodes"=>[:color, :spin]))
\end{lstlisting}

In the step rule of the Ising model as shown in listing~\ref{lst:Isingstep}, we carry out a fixed number of Monte-Carlo steps in which a node is selected at random and its spin and color are flipped depending on the usual Ising energy condition. Then we run the model for 100 steps using the \verb+run_model!+ function and sending \verb+step_rule!+ as an argument. 

\begin{lstlisting}[caption = {Ising model on nearest neighbor graph - step run.}, label={lst:Isingstep}]
function step_rule!(model)
	nodesprops = model.graph.nodesprops
    for i in 1:100
        rand_nd = rand(1:n)
        spin = nodesprops[random_node].spin
        nbr_nodes = neighbor_nodes(rand_nd, model)
        de = 0.0
        for node in nbr_nodes
            nbr_spin = nodesprops[node].spin
            de += spin*nbr_spin
        end
        de = 2*model.parameters.coupl * de
        if (de < 0) || (rand() < exp(-de/model.parameters.temp))
            nodesprops[random_node].spin = - spin
            nodesprops[random_node].color = spin == -1 ? :black : :white
        end
    end
end

run_model!(model, steps = 100, step_rule = step_rule!)
\end{lstlisting}

\begin{figure*}[tb]
\centering
\includegraphics[width=\textwidth]{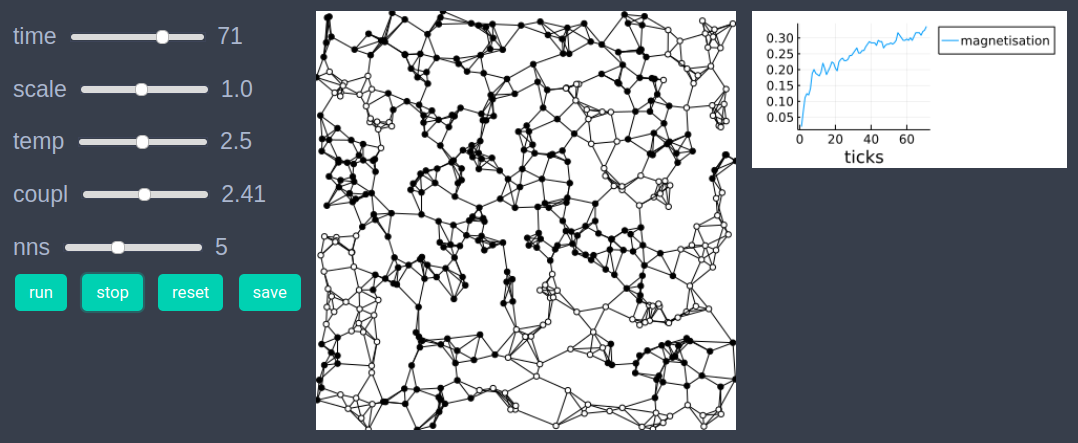}
\caption{\label{fig:isingapp}Interactive app for Ising model on a nearest neighbor graph. Sliders can be used to change the model paramters and then the model can be re-initialised with the reset button.}
\end{figure*}

Tasks of creating an animation and interactive app for graph models uses the same functions as in 2D and 3D case. For example, the code below will create an interactive app for the Ising model as shown in Fig~\ref{fig:isingapp}

\begin{lstlisting}
create_interactive_app(model, 
	initialiser= initialiser!,
    props_to_record = 
    Dict("nodes"=>[:color, :spin]),
    step_rule= step_rule!,
    model_controls=[
    (:temp, :s, 0.05:0.05:5.0),
    (:coupl, :s, 0.01:0.1:5.0),
    (:nns, :s, 2:10)],
    node_plots = Dict("magnetisation"=> 
    			x -> x.spin),
    frames=100) 
\end{lstlisting}

\begin{figure*}[tb]
\centering
\includegraphics[scale=0.4]{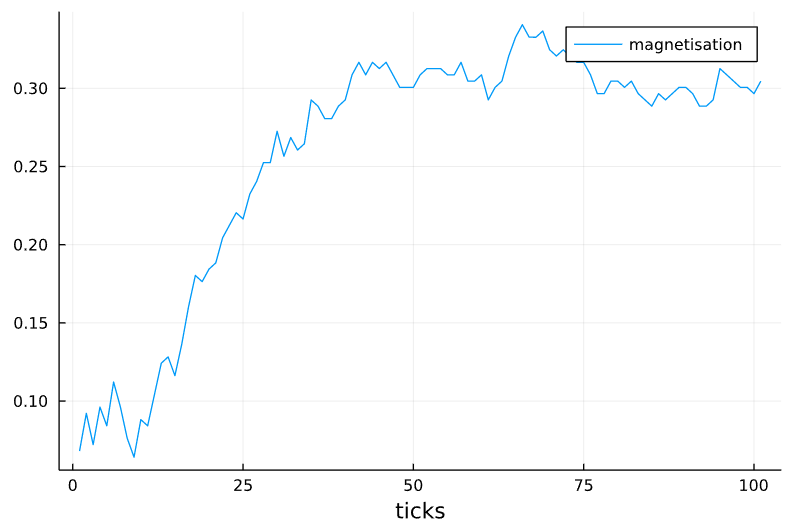}
\caption{\label{fig:isingplot}Plot of magnetisation vs time for the Ising model.}
\end{figure*}

EasyABM.jl API provides simple functions for fetching data of individual agents, nodes and edges as in case of 2D and 3D models. For example, the code below fetches the data of average spin of nodes (also called magnetisation) as a dataframe and plots the result as shown in Fig~\ref{fig:isingplot}.

\begin{lstlisting}
df = get_nodes_avg_props(model, 
		node -> node.spin, 
		labels=["magnetisation"], 
		plot_result = true)
\end{lstlisting}

\section{Conclusions and future work}
We have introduced a new Julia framework for agents based modelling and explained its workflow with examples of 2D, 3D and graph based models. Some of the notable features of EasyABM we have tried to emphasise are -
\begin{itemize}
\item Fully function based approach. 
\item Intuitive visualisation.  
\item Easy data collection. 
\item Patches of space, and nodes and edges of graph can be assigned properties like agents. 
\item Dynamic graph spaces.  
\end{itemize}

Some of the future plans for EasyABM are - 
\begin{itemize}
\item Inclusion of more inbuilt graph types (current version only has some inbuilt grid graphs).
\item Providing more than one backend options for visualisations in 2D, 3D and graph models. 
\item Visualisation of graphs in 3D space. 
\item Expanding documentation by including more examples. 
\end{itemize}

Users of EasyABM are also welcome to contribute to the package by suggesting or adding new features, improving documentation and reporting bugs.

\clearpage
\balance


\begin{thebibliography}{10}

\bibitem{Nic2020}
Nicolas Hoertel, Martin Blachier, Carlos Blanco, Mark Olfson, Marc Massetti, Marina Sánchez Rico, Fr\'{e}d\'{e}ric Limosin and Henri Leleu, 
\textit{A stochastic agent-based model of the SARS-CoV-2 epidemic in France}, Nature Medicine, vol.~26, pp.~1417--1421, 2020.

\bibitem{Hinch2021}
Hinch R, Probert WJM, Nurtay A, Kendall M, Wymant C, Hall M, et al., \textit{OpenABM-Covid19-An agent-based model for non-pharmaceutical interventions against COVID-19 including contact tracing}, PLoS Computational Biology, vol.~17, issue~7, 2021.

\bibitem{Kerr2021}
 Kerr CC, Stuart RM, Mistry D, Abeysuriya RG, Rosenfeld K, Hart GR, et al., \textit{Covasim: An agent-based model of COVID-19 dynamics and interventions}, PLoS Computational Biology, vol.~17, issue~7, 2021.
 
 \bibitem{SM2014}
 SM Niaz Arifin, Ying Zhou, Gregory J.~Davis, James E.~Gentile, Gregory R.~Madey and Frank H.~Collins, \textit{An agent-based model of the population dynamics of Anopheles gambiae}, Malaria Journal, vol.~13, Article number 424, 2014. 
 
 \bibitem{Jang2016}
 Jang Won Bae, Euihyun Paik. Kiho Kim, Karandeep Singh and Mazhar Sajjad,
 \textit{Combining Microsimulation and Agent-based Model for Micro-level Population Dynamics}, Procedia Computer Science, vol.~80, pp.~507--517, 2016.
 
 \bibitem{Lardon2011}
 Lardon LA, Merkey BV, Martins S, D\"{o}tsch A, Picioreanu C, Kreft JU, Smets BF, \textit{iDynoMiCS: next-generation individual-based modelling of biofilms}, Environmental Microbiology. vol.~13, issue~9, pp.~2416--2434, September 2011.
 
 \bibitem{Jan2016}
 Jan Poleszczuk, Paul Macklin and Heiko Enderling,
 \textit{Agent-Based Modeling of Cancer Stem Cell Driven Solid Tumor Growth}, Methods in Molecular Biology, vol.~1516, pp.~335--346, 2016.
 
 \bibitem{Sam2020}
 Samuel Vanfossan, Cihan H.~Dagli and Benjamin Kwasa, 
 \textit{An Agent-Based Approach to Artificial Stock Market Modeling}, Procedia Computer Science, vol.~168, pp.~161--169, 2020. 
 
 \bibitem{Feng2012}
 Ling Feng, Baowen Li, Boris Podobnik, Tobias Preis, and H.~Eugene Stanley, 
 \textit{Linking agent-based models and stochastic models of financial markets}, Applied Physical Science, PNAS, vol.~109, issue~22, May 2012.
 
 \bibitem{Hager2015}
 Karsten Hager, J\"{u}rgen Rauh and Wolfgang Rid,
 \textit{Agent-based Modeling of Traffic Behavior in Growing Metropolitan Areas}, Transportation Research Procedia, vol.~10, pp.~306--315, 2015.
 
 \bibitem{Mau2021}
 Mauricio Gonz\'{a}lez-M\'{e}ndez, Camilo Olaya, Isidoro Fasolino, Michele Grimaldi and Nelson Obreg\'{o}n, 
 \textit{Agent-Based Modeling for Urban Development Planning based on Human Needs. Conceptual Basis and Model Formulation}, Land Use Policy, vol.~101, Feb 2021. 
 
  \bibitem{Anna2016}
 Anna Klabunde and Frans Willekens,
 \textit{Decision-Making in Agent-Based Models of Migration: State of the Art and Challenges}, European Journal of Population, vol.~32, pp.~73--97, Feb 2016. 
 
 \bibitem{Jule2018}
 Jule Thober, Nina Schwarz and Kathleen Hermans,
 \textit{Agent-based modeling of environment-migration linkages: a review}, Ecology and Society, vol.~23, issue~2, June 2018.

\bibitem{Wilensky1999}
L.~Wilensky, \textit{Netlogo}, Center for Connected Learning and Computer-Based Modeling, Northwestern University, 1999.

\bibitem{SLuk2005}
S.~Luke, C.~Cioffi-Revilla, L.~Panait, K.~Sullivan, and G.~Balan,
\textit{MASON: A Multiagent Simulation Environment}, SIMULATION,
 vol.~81, pp.~517--527, July 2005.
 
\bibitem{MJ2013}
M.~J. North, N.~T. Collier, J.~Ozik, E.~R. Tatara, C.~M. Macal, M.~Bragen, and P.~Sydelko, \textit{Complex adaptive systems modeling with Repast Simphony}, Complex Adaptive Systems Modeling, vol.~1, p.~3, Dec. 2013.

\bibitem{HIba2013}
H.~Iba, \textit{Agent-Based Modeling and Simulation with Swarm},
Chapman and Hall/CRC, zeroth~ed., June 2013.


\bibitem{DMas2015}
D.~Masad and J.~Kazil, \textit{Mesa: An Agent-Based Modeling Framework}, Python in Science Conference, pp.~51--58, 2015.

\bibitem{George2022}
George Datseris and Ali~R. Vahdati and Timothy~C. DuBois \textit{Agents.jl: a performant and feature-full agent-based modeling software of minimal code complexity}, Simulation, Sage Journals, 2022.

\bibitem{EasyABM}
Renu Solanki, Monisha Khanna Kapur,
Shailly Anand, Anita Gulati, 
Prateek Kumar, Munendra Kumar and Dushyant Kumar,
\textit{EasyABM.jl github repository
  documentation}. \url{}.


\end{thebibliography}
\end{document}